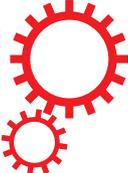

# Efficient artificial mineralization route to decontaminate Arsenic(III) polluted water - the Tooeleite Way

Arindam Malakar[1], Bidisa Das[2], Samirul Islam[1], Carlo Meneghini[3], Giovanni De Giudici[4], Marco Merlini[5], Yury V. Kolen'ko[6], Antonella Iadecola[7], Giuliana Aquilanti[7], Somobrata Acharya[2] & Sugata Ray[1,2]

Increasing exposure to arsenic (As) contaminated ground water is a great threat to humanity. Suitable technology for As immobilization and removal from water, especially for As(III) than As(V), is not available yet. However, it is known that As(III) is more toxic than As(V) and most groundwater aquifers, particularly the Gangetic basin in India, is alarmingly contaminated with it. In search of a viable solution here, we took a cue from the natural mineralization of Tooeleite, a mineral containing Fe(III) and As(III) ions, grown under acidic condition, in presence of $SO_4^{2-}$ ions. Complying to this natural process, we could grow and separate Tooeleite-like templates from Fe(III) and As(III) containing water at overall *circumneutral* pH and *in absence of* $SO_4^{2-}$ ions by using highly polar Zn-only ends of wurtzite ZnS nanorods as insoluble nano-acidic-surfaces. The central idea here is to exploit these insoluble nano-acidic-surfaces (called as INAS in the manuscript) as nucleation centres for Tooeleite growth while keeping the overall pH of the aqueous media neutral. Therefore, we propose a novel method of artificial mineralization of As(III) by mimicking a natural process at nanoscale.

Arsenic, above a certain threshold level, is extremely toxic for human body[1–4]. Arsenic bearing minerals are abundant in the Earth's crust which can gradually dissolve in groundwater from weathered rocks and soils[1,5], thereby increasing contamination and toxicity in places with no history of arsenic related health problems[6,7]. Most disturbingly, human activities like mining and industrialization can further aggravate arsenic contamination locally. Therefore, the search for efficient method(s) of stable and long-serving arsenic remediation is essential.

The common arsenic species in circumneutral natural water are hydrated anions of As(V) ($H_2AsO_4^-$ and $HAsO_4^{2-}$), and the hydrated As(III) - the neutral molecule $H_3AsO_3$. Existing arsenic removal techniques[1,8–12] generally rely on surface adsorption of charged contaminant ions on adsorbent surfaces which fail to work for neutral $H_3AsO_3$[1,13–17]. Therefore, there exists no single efficient method of direct As(III) removal. In contrary, As(III) is significantly more toxic than As(V)[18], and more mobile in water[17] which makes it one of the greatest dangers to humanity. For example, in the Gangetic basin of Indian peninsula, almost 100 million people are currently at a risk of As(III) contamination of varying degrees[19–21]. At present, As(III) remediation employs pre-oxidation of As(III) to As(V) and subsequent removal of As(V) by standard adsorption-based techniques[1,8,9]. Even then these efforts have serious limitations because adsorption-based technologies do not permanently remove As(V); which may simply be re-introduced into the natural cycle under reducing conditions[22].

Interestingly, a few viable solutions to such contamination problems involving certain microorganisms (biomineralization) already exist in nature. For example, the oxidation of Fe(II) in acid mine drainage water containing both Fe(II) and As(III) results in trapping of As(III) by *Acidithiobacillus ferrooxidans* strains into Tooeleite mineral[23–26] (general formula $Fe_6(AsO_3)_4SO_4(OH)_4\cdot4H_2O$). Such mineralization processes strongly limit the mobility of heavy metal ions such as arsenic and could offer an effective remediation strategy. However, natural crystalline Tooeleite forms in acidic medium, with $SO_4^{2-}$ ions and at prolonged timescales[23–27]. Therefore, natural Tooeleite formation process as such cannot be conceived as an As(III) removal method from natural water, unless there are means to implement it at neutral pH and without $SO_4^{2-}$ ions (usually absent in natural groundwater). Here, we show that it is possible to instigate nucleation of Tooeleite-like template even in natural contaminated water (pH~7) devoid of $SO_4^{2-}$ ions, bearing both As(III) and Fe(III) (Fe(III) is also a common natural contaminant) with the help of wurtzite ZnS nanorods. The underlying strategy here is to provide large number of insoluble nano-acidic-surfaces (will be abbreviated as INAS from now on) in the form of Zn-only polar planes of ZnS nanorods[28–30] inside water, which initiates Tooeleite-like template formation within otherwise neutral aqueous media under moderate heating for 3 hours. This mechanism is unequivocally confirmed by our theoretical modelling and finally, we are being able to suggest a hitherto unknown efficient pathway of direct As(III) remediation of water.

[1]Department of Materials Science, Indian Association for the Cultivation of Science, Jadavpur, Kolkata 700032, India. [2]Centre for Advanced Materials, Indian Association for the Cultivation of Science, Jadavpur, Kolkata 700032, India. [3]Dipartimento di Scienze, Universitá Roma Tre, Via della Vasca Navale, 84 I-00146 Roma, Italy. [4]Department of Chemical and Geological Sciences, University of Cagliari, 09127 Cagliari, Italy. [5]Universita di Milano Dip. di, Scienzedella Terra Ardito Desio, Milano, Italy. [6]International Iberian Nanotechnology Laboratory, Av. Mestre José Veiga s/n, 4715-330 Braga, Portugal. [7]Elettra-Sincrotrone Trieste S.C.p.A., Strada Statale 14, km 163.5, 34149 Basovizza, Trieste, Italy. Correspondence and requests for materials should be addressed to S.R. (email: mssr@iacs.res.in)

## Results and Discussion

The wurtzite structure possesses extremely polar all-cationic and all-anionic crystallographic planes and research shows[28,29] that it tends to grow in a cylindrical rod-like morphology with nonpolar lateral surfaces and high surface energy polar bases consisting of all $Zn^{2+}$ (0001) or all $S^{2-}$ (000$\bar{1}$) ions. A representative crystal structure of such a nanorod is shown in Fig. 1(a). We have previously demonstrated that polar planes, normal to the long axis of the nanorods, can trigger many unusual reactions/phase formations in water[30]. Here, we exploit this unique feature of ZnS nanorods for As(III) remediation, where $Zn^{2+}$ (0001) polar planes have been supplied in water as INAS. High-resolution TEM images of the as-grown nanorods are shown in Fig. 1(b), indicating the presence of reasonably monodisperse nanorods 8–10 nm in length and 1–1.5 nm in diameter. The Zn-only ends of these nanorods are highly polar and of Lewis acidic nature (INAS), which floats around in neutral water offering a conducive atmosphere to the Fe(III) and As(III) ions to get mineralized in form of Tooeleite.

In this work, we experimented with three different samples of water, i.e., As(III) and Fe(III) contaminated ground water from West Bengal, India (NW-Kolkata: 'NW' denotes natural water), a laboratory-made aqueous solution of $Fe^{3+}$ and $As^{3+}$ cations intended to mimic the concentrations of the previous sample (AF-Lab: 'AF' denotes As and Fe), and As-spiked natural water from Sardinia island, Italy ('NW'-Sardinia: see Supplementary Table S1). The first two samples had ~0.4 mg/L of arsenic, and the sample from Sardinia contained ~0.2 mg/L (Supplementary Table S1); the permissible amount of arsenic in potable water as defined by the World Health Organization is 0.01 mg/L[31]. Notably, the arsenic concentration in natural water sources vary quite significantly at different times of the year, especially before and after the rainy season (Supplementary Table S1)[5]. We thus collected water from the local area at different times of the year to confirm the efficiency and consistency of our method (see inset of Fig. 1(c)). However, in this paper, we only present detailed results from samples with the highest initial arsenic content. The simple treatment that we propose here is to add a given quantity of ODA-capped ZnS nanorods to a given volume of sample water and then heat the solution at 70 °C for a duration of 3 hours under continuous stirring. For analysis, the resultant solutions were filtered using Whatman 42 filter paper (~2.5 μm pore size), and the corresponding filtrates and residues (whenever available) were collected. In all cases, the solutions turned yellow with turbidity (right lower inset to Fig. 1(d)) after treatment. All samples (samples NW-Kolkata_ZnS-NR, AF-Lab_ZnS-NR and NW-Sardinia_ZnS-NR) before and after treatment were quantitatively analysed using ICP-OES/MS following standard protocol for determination of As and Fe concentrations. Water samples without ZnS nanorods were also treated identically, which are 'Blank' samples (NW-Kolkata_Blank, AF-Lab_Blank, and NW-Sardinia_Blank) and analysed quantitatively before and after the heat treatment.

The experimentally obtained removal percentages of As and Fe are summarized in Fig. 1(c). Figure 1(c) reveals that treatment with ZnS nanorods works efficiently and removes more than 90% of the contaminant As(III) ions, whereas the 'Blank' treatment results in $\leq$10% removal. The NW-Kolkata sample is expected to contain some As(V) along with large amounts of As(III), as is known from arsenic speciation studies of the region[32–36], which is likely the reason for ~30% arsenic removal, even without ZnS nanorods (see NW-Kolkata_Blank in Fig. 1(c)). However, the AF-Lab and NW-Sardinia samples consisted of only As(III) by design, and so it can be concluded that this single-step treatment removes 90±5% of total As(III). The inset to Fig. 1(c) shows the consistency of our method, where samples collected year-round in Kolkata were treated with ZnS nanorods and it is observed that arsenic is reduced to <10 ppb in most of the cases. Even for samples, which have a relatively small initial As-content (~50 ppb), the contaminant is reduced convincingly below 10 ppb, which is otherwise known to be a daunting task[1]. One important point to note here that our method is able to remove nearly 7g As/g of ODA-coated ZnS, which in an average is three orders of magnitude more efficient than other known methods[9,15,37–39] of As(III) removal. Moreover, nearly comparable removal efficiency observed with natural water samples (NW-Kolkata and NW-Sardinia) as well as laboratory water (AF-Lab) confirms that other ions do not adversely affect the reaction to any noticeable extent. Fig. 1(d) provides a representative TEM image of the residual solid of NW-Kolkata_ZnS-NR sample and in each case, spindle-like microparticles (~1 μm in length) are found, which readily precipitates and is removed by standard pore-filtration. High Resolution TEM (right upper inset of Fig. 1(d)) shows definite lattice patterns and in selected area electron diffraction (SAED) (left inset of Fig. 1(d)) spots are clearly visible which indicates the residue is crystalline in nature.

The residual microparticles were characterized in details to identify the As-containing phase(s) and to understand the chemical mechanism of the phase formation. Figure 2(b,d,f) provides 2D-XRD images from AF-Lab_ZnS-NR, NW-Kolkata_ZnS-NR, and NW-Sardinia_ZnS-NR residues, and the extracted θ–2θ patterns are shown in Fig. 2(a,c,e). It was observed that XRD pattern of the AF-Lab_ZnS-NR residue is similar to that of Tooeleite[23–26,40,41] ($Fe_6(AsO_3)_4SO_4(OH)_4.4H_2O$)), with certain differences, as $SO_4^{2-}$ is absent in AF-Lab_ZnS-NR residue. Moreover, the pH of the solution was maintained neutral throughout the reaction, unlike acidic medium required for natural Tooeleite[23–27]. The situation is more complex for natural water samples with multiple ions and microorganisms, where various crystalline phases are formed. The difficulty in analyzing the natural water residues is clear from Fig. 2(d,f), where with weak Tooeleite-like diffraction rings, intense spots of $Ca^{2+}$ bearing Calcite phase and broader rings of Aragonite phase can be identified. However, the high statistics/resolution patterns obtained from SR-XRD allowed for a multiphase full-profile Rietveld refinement analysis and reliable recognition of the Tooeleite pattern (Fig. 2(c)). The phase identification was more difficult for NW-Sardinia_ZnS-NR residue (Fig. 2(e,f)), where Tooeleite-like peaks were hidden even in the extracted pattern. Nevertheless, it was still possible to identify the presence of a Tooeleite-like structure (see enlarged version of Fig. 2(e)) as the As-bearing entity. Rietveld refinements of these data (except for the NW-Sardinia_ZnS-NR residue) were performed, and structural parameters were extracted (Supplementary Table S2). These results were further confirmed by As $K$-edge Extended X-ray Absorption Fine Structure (EXAFS) spectra (shown in Fig. 2(g)) from which all the local atomic distances were obtained and compared with existing Tooeleite data[26] (see Table 1). As the formation of Tooeleite-like structure is confirmed in every case, one also understands the reason for the development of yellow color in the resultant solution every time, which is indeed precipitation of the yellow colored mineral Tooeleite during reaction.

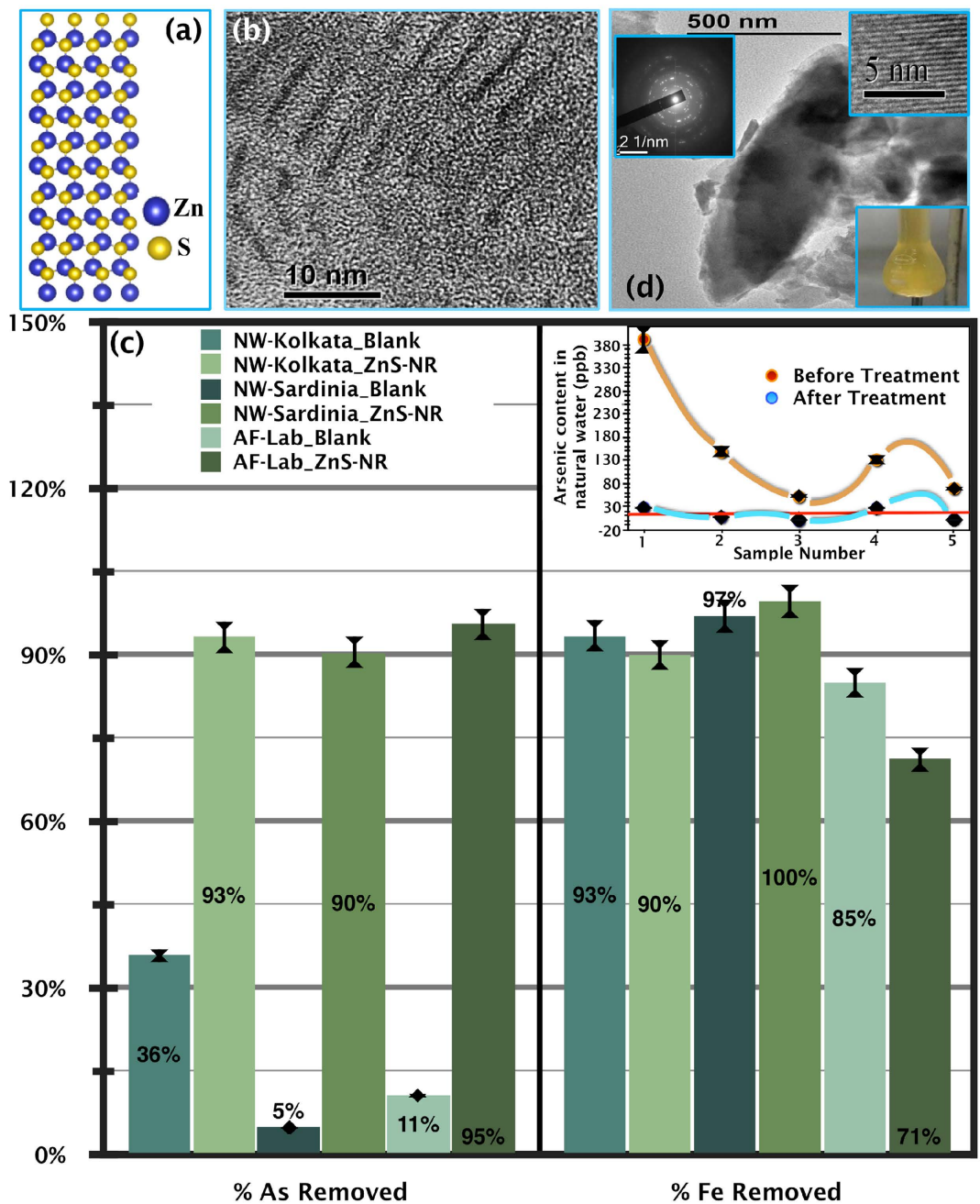

**Figure 1.** (**a**) Model of as synthesized ZnS nanorods, (**b**) TEM image of as-synthesized ZnS nanorods, (**c**) Percentage removal of arsenic and iron from natural water (from Kolkata, India (NW-Kolkata) and Sardinia, Italy (NW-Sardinia)) and artificial water (equivalent to the concentration of natural water of Kolkata (AF-Lab)) in the absence (Blank) and presence of ZnS nanorods (ZnS-NR). Inset shows the arsenic concentration (in ppb) in natural water collected during different time periods and locations in India before (orange dot) and after (blue dot) treatment with ZnSnanorods; World Health Organization (WHO) guideline of 10 ppb (red line) is provided for comparison. (**d**) Shows TEM image of as-formed microcrystals from sample NW_Kolkata-ZnS_NR. Inset shows HRTEM of the same microcrystals, SAED diffraction of the microcrystal and the image of the actual product (light yellowish coloured) formed after ZnS treatment of NW-Kolkata.

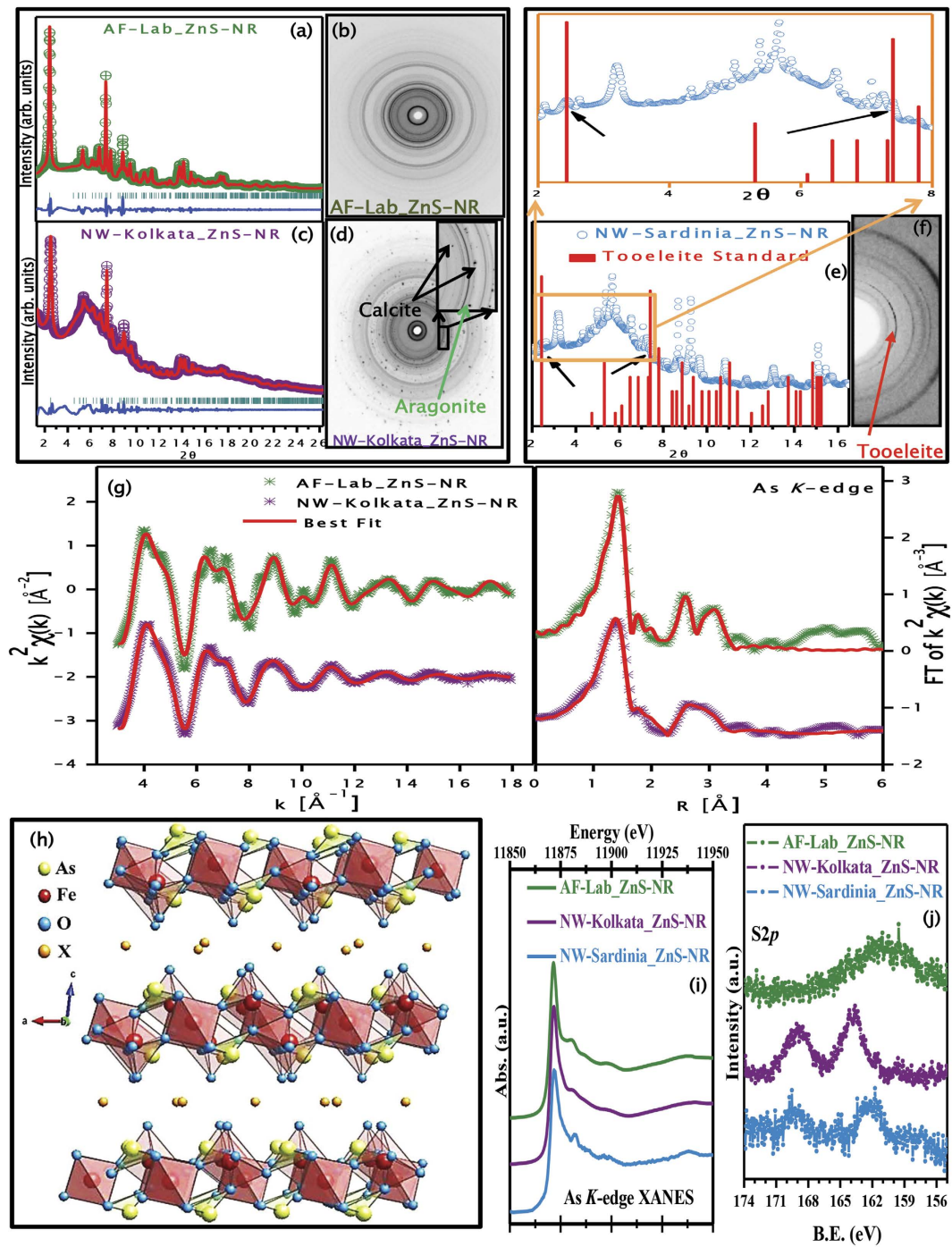

**Figure 2.** Deconvoluted powder XRD pattern obtained from 2D-XRD of samples (**a**) AF-Lab_ZnS-NR, (**c**) NW-Kolkata_ZnS-NR, and (**e**) NW-Sardinia_ZnS-NR. Actual 2D-XRD images of samples (**b**) AF-Lab_ZnS-NR, (**d**) NW-Kolkata_ZnS-NR (along with other phases from calcium containing minerals, such as calcite and aragonite), and (**f**) NW-Sardinia_ZnS-NR (contains multiple new phases). (**g**) As $K$-edge (experimental and best fit (red line)) obtained for AF-Lab_ZnS-NR (green) and NW-Kolkata_ZnS-NR (purple). (**h**) Crystal structure obtained by Rietveld refinement of deconvoluted XRD pattern from AF-Lab_ZnS-NR, (**i**) As $K$-edge XANES spectra and (**j**) Photoemission spectra of S $2p$ for different samples.

| Bond Type | As-O | | As-Fe1 | | As-Fe2 | |
|---|---|---|---|---|---|---|
| Sample | $N_i$ | $R_i$ (Å) (1.797) | $N_i$ | $R_i$ (Å) (2.889) | $N_i$ | $R_i$ (Å) (3.485) |
| NW-Kolkata_ZnS-NR | 3 | 1.783(3) | 1 | 2.956(1) | 3 | 3.487(5) |
| AF-Lab_ZNS-NR | 3 | 1.784(3) | 1 | 2.905(1) | 3 | 3.499(5) |
| Bond Type | Fe-O | | Fe1-As | | Fe2-As | | Fe1-Fe2 | |
| Sample | $N_i$ | $R_i$ (Å) (2.036) | $N_i$ | $R_i$ (Å) (2.889) | $N_i$ | $R_i$ (Å) (3.459) | $N_i$ | $R_i$ (Å) (3.589) |
| NW-Kolkata_ZnS-NR | 6 | 1.983(2) | 0.33 | 2.95(2) | 2.66 | 3.40(6) | 4 | 3.58(5) |
| AF-Lab_ZNS-NR | 6 | 1.985(2) | 0.33 | 2.90(2) | 2.66 | 3.42(6) | 4 | 3.59(5) |

**Table 1. Coordination numbers and bond lengths obtained from EXAFS analysis of samples.** Coordination numbers were fixed according to the Tooeleite[26] crystallographic structure (ICSD code 156179), whereas interatomic distances are refined; the standard deviation over the 8 samples analyzed is reported in parentheses. The interatomic distances of the ideal Tooeleite crystallographic structure are shown in bold.

The diffraction data from sample AF-Lab_ZnS-NR allowed for *ab-initio* structure determinations using the charge-flipping algorithm approach, and the model confirmed the presence of the Fe-As-O layers closely resembling Tooeleite, as shown in Fig. 2(h). The central template is nearly identical to that of Tooeleite; made by corner linked Fe octahedra stapled by As in a bridging configuration. These templates are separated by non-bonded anionic entities, which are shown as orange spheres in the image. Furthermore, to probe whether As in the residue were in +3 or +5 oxidation state(s), core level X-ray photoelectron spectroscopy (XPS) and X-ray absorption near edge spectroscopy (XANES; As and Fe *K*-edge) measurements on all samples were carried out. Whereas the Fe 2*p* XPS and *K*-edge XANES spectra (see Supplementary Fig. S2(a,b)) proved +3 oxidation state, the As 3*d* XPS spectra (Supplementary Fig. S2(c) available online) and As *K*-edge XANES spectra (Fig. 2(i)) confirmed the presence of As in +3 states in all residual solids. These results establish that indeed synthesized structures mimic the central template of the Tooeleite mineral, chemically and structurally. There could still be arguments regarding the presence and form of sulphur in the residues as externally added ZnS may act as a source of sulphur (although only 20% of the total sulphur needed to form the ideal Tooeleite composition for the amount of residue obtained is provided by the added ZnS) and may even provide $SO_4^{2-}$ anions. To further confirm this point, sulphur 2*p* core-level XPS experiments were carried out (see Fig. 2(j)), to identify different sulphur anions[42]. It is known that S 2*p* core levels for $S^{2-}$ and $SO_4^{2-}$ anions appear at two different binding energies[42–46], (161.5 eV, as in ZnS and 169.0 eV, as in Tooeleite respectively) and in our experiments, the major sulphur component is $S^{2-}$, though there is also an accompanying $SO_4^{2-}$ signal in some cases. Moreover, in AF-Lab_ZnS-NR where the only source of sulphur could be from the added ZnS nanorods, only the $S^{2-}$ signal is observed (see Fig. 2(j)). Therefore, it is established from all the chemical and structural characterizations that our procedure of '*artificial mineralization*' allows As(III) removal and incorporation on a mineral phase within a structure closely resembling Tooeleite, made up of As-Fe-O layers, separated by non-bonded molecules (e.g. water).

Next, quantum chemical methods were employed to understand the formation of the Tooeleite-like central template in water bearing Fe(III) and As(III) in presence of ZnS nanorods. Generally, Fe(III) ions in water exists as $Fe(H_2O)_6$ (**A**, Fig. 3(a)), which may spontaneously dimerise to form an oxyhydroxide ($Fe_2(H_2O)_6(OH)_4^{2+}$: **B**), with two -OH bridging two Fe-centres (Fig. 3(b))[47]. Aqua linkages too are possible, but hydroxy linkages are more stable. The enthalpy of formation of **A** and **B** are,

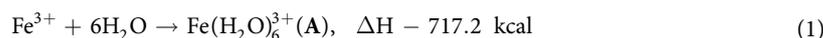
$$Fe^{3+} + 6H_2O \rightarrow Fe(H_2O)_6^{3+} (\mathbf{A}), \quad \Delta H - 717.2 \text{ kcal} \quad (1)$$

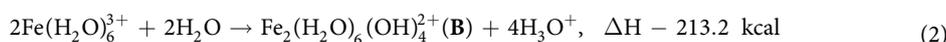
$$2Fe(H_2O)_6^{3+} + 2H_2O \rightarrow Fe_2(H_2O)_6(OH)_4^{2+} (\mathbf{B}) + 4H_3O^+, \quad \Delta H - 213.2 \text{ kcal} \quad (2)$$

In case of complex **B** there are two OH bridges between the two central Fe atoms, and Fe-Fe distance is 2.89 Å and Fe-O(bridge) distance is 1.86 Å, whereas other Fe-O (O from bound water molecules) distances range from 2.01–2.04 Å (Fig. 3(b)). Therefore, it can be assumed that in experimental conditions, complexes such as **A**, **B** along with neutral $H_3AsO_3$ are present in the water medium to which ZnS nanorods are added. It is observed that hydrated $Fe(H_2O)_6^{3+}$, $As(OH)_3$ and Fe-dimer **B** may form stable inorganic complex **BFe$_4$As$_1$** (Fig. 3(c)), as

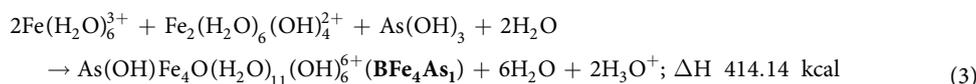
$$2Fe(H_2O)_6^{3+} + Fe_2(H_2O)_6(OH)_4^{2+} + As(OH)_3 + 2H_2O$$
$$\rightarrow As(OH)Fe_4O(H_2O)_{11}(OH)_6^{6+}(\mathbf{BFe_4As_1}) + 6H_2O + 2H_3O^+; \Delta H \ 414.14 \text{ kcal} \quad (3)$$

In the complex **BFe$_4$As$_1$**, the distance between two Fe centers is 2.98 Å and the Fe-O distances are 1.90 Å, while the Fe-As distance is 2.99 Å, As-O distances are 1.76–1.86 Å and Fe-O distances range from 1.85–1.95 Å. Since the enthalpy of formation is positive, the reaction is not spontaneous. Once **BFe$_4$As$_1$** is formed, the reaction with more molecules of $As(OH)_3$ are favorable and formation of **BFe$_4$As$_4$** (Tooeleite, Fig. 3(d)) is feasible,

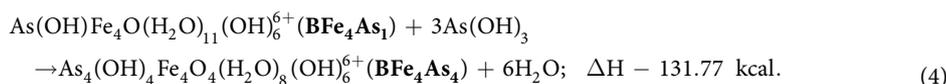
$$As(OH)Fe_4O(H_2O)_{11}(OH)_6^{6+}(\mathbf{BFe_4As_1}) + 3As(OH)_3$$
$$\rightarrow As_4(OH)_4Fe_4O_4(H_2O)_8(OH)_6^{6+}(\mathbf{BFe_4As_4}) + 6H_2O; \quad \Delta H - 131.77 \text{ kcal}. \quad (4)$$

To explore the role of ZnS nanorods in forming the tooeleite template and to understand the experimental observations, we theoretically try to model the formation of **BFe$_4$As$_1$** in presence of ZnS nanostructures. As a model nanorod we have used a small cluster of $Zn_{14}S_{14}$, which mimicks the wurtzite nature of the ZnS nanorod. The uniqueness of this nanorod structure is its polar ends, i.e it has one pure Zn end and one pure S end which are highly reactive. In the studies involving model $Zn_{14}S_{14}$ cluster a partial optimization of atomic positons were carried out, where the positions of only first layer Zn atoms directly interacting with $As(OH)_3$ are allowed to change and the rest three layers were kept fixed[30]. In any such adsorption reaction the surface atoms are most reactive, thus optimising only the first layer is not very unrealistic. After optimization of the $Zn_{14}S_{14}$ cluster when the positions of the top Zn atoms were relaxed, the surface Zn-Zn bond-distances decrease, as shown in Fig. 3(e). To study the formation of **BFe$_4$As$_1$** from its reactants in the presence of $Zn_{14}S_{14}$, we probed the simultaneous formation and subsequent binding of **BFe$_4$As$_1$** to $Zn_{14}S_{14}$ cluster (Fig. 3(f), where the oxygen atoms connected to As are bound to the Zn layer) from the reaction:

$$Zn_{14}S_{14} + 2Fe(H_2O)_6^{3+}(\mathbf{A}) + Fe_2(H_2O)_6(OH)_4^{2+}(\mathbf{B})$$
$$+ As(OH)_3 \rightarrow Zn_{14}S_{14} - \mathbf{BFe_4As_1}^{4+} + 4H_3O^+ \tag{5}$$

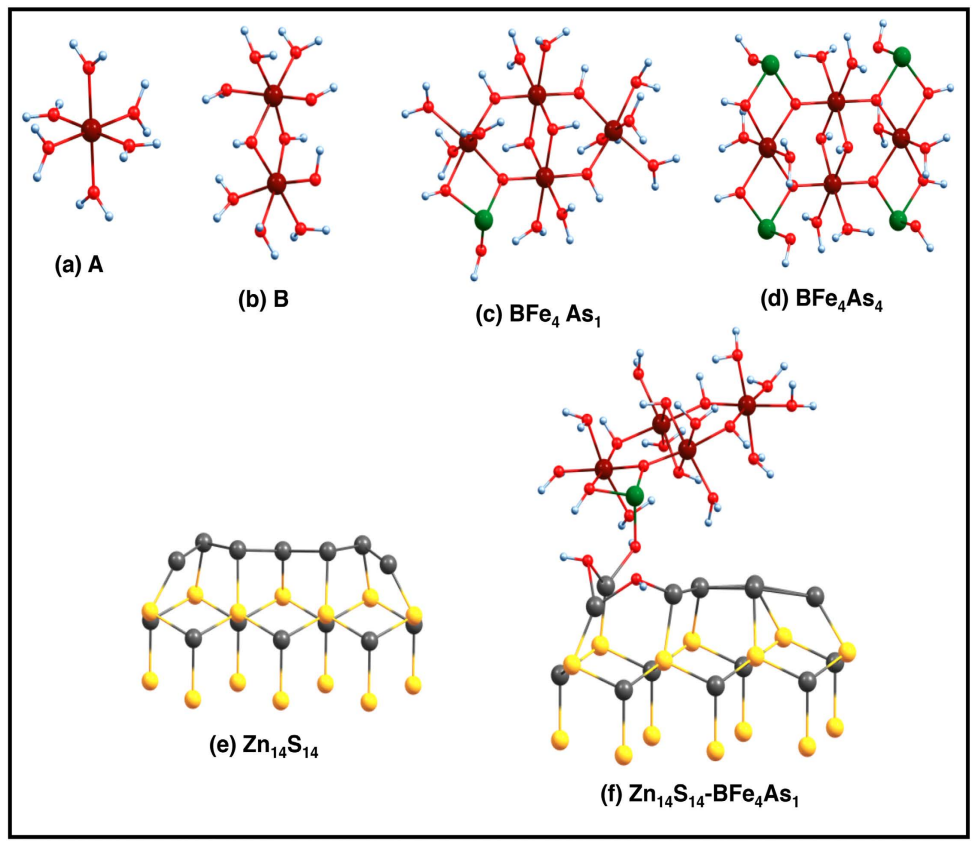

**Figure 3. Optimized structures of various species involved are shown in ball and stick models.** The colour code is brown: Fe(III), red: O, green: As(III), grey: $Zn^{2+}$, yellow: $S^{2-}$ and small bluish-white spheres are hydrogens.

We find a large binding energy (~1299 kcal/mol), indicating the spontaneous formation of **BFe$_4$As$_1$** complex bound to $Zn_{14}S_{14}$ cluster. The high binding energy stems from strong binding of $As(OH)_3$ on the free Zn surface of the ZnS nanorod. It is seen that, for the chemisorption reaction $Zn_{14}S_{14} + As(OH)_3 \rightarrow (Zn_{14}S_{14})$-$As(OH)_3$ (chemisorbed product), the binding energy is:

$$E_{binding} = E\_[(Zn_{14}S_{14}) - As(OH)_3] - [E\_(Zn_{14}S_{14}) + E\_As(OH)_3],$$
$$E_{binding} = 1108.65 \text{ kcal} \tag{6}$$

Close inspection shows, upon adsorption of $As(OH)_3$ on Zn surface, As-OH distances increase from 1.8 Å as in $As(OH)_3$, only one OH group is retained solely by As and two OH groups interact less with As forming bridge bonds to free surface Zn atoms. Thus, it could be concluded that the reactive free Zn end of the ZnS nanorods does act as INAS and bind to the -OH groups of $As(OH)_3$ virtually causing an elongation of As-OH bonds which makes the As(III) ions more available for attachment to the Fe-complex, and the $Zn_{14}S_{14}$-**BFe$_4$As$_1$** moiety is easily formed (Fig. 3(f)).

Overall, we show that treatment of contaminated water bearing As(III) and Fe(III) ions with ZnS nanorods at 70 °C for 3 hours permanently removes the contaminants. Our process has two significant improvements over the natural one; it works at neutral pH without $SO_4^{2-}$ ions, and the reaction time is shorter (3 hours). This is achieved by supplying INAS in water in the form of Zn-only ends of the ZnS nanorods instead of a bulk acidic environment as is needed for natural growth of Tooeleite. Moreover, ZnS nanorods break completely after treatment, so no secondary nanoparticle based pollutant is produced. We believe that our results present an archetype to entrap As(III) in artificial minerals and many permanent remedies of environmental contamination issues can be addressed through such *artificial mineralization* processes, and current research should be focused along these directions.

## Methods

**Materials.** ODA (Sigma, >99%), potassium ethyl xanthogenate (Sigma, >96%), zinc acetate dihydrate (Sigma, 98%) were used as received. The ODA was stored in argon all the time prior to use. Ethanol absolute (VWR International, Normapur) and Chloroform (Emparta ACS) were used for synthesis. KOH (Chem Labs, 85%), $As_2O_3$ (Nice Laboratories ~99.5%) and Suprapur $HNO_3$ (Merck, Germany, 65%) was used to make AF-Lab water. Deionized water (resistivity 18.2 MΩ cm) used for making artificial arsenic contaminated water, was obtained from a Millipore filter system.

**Nanoparticle Synthesis.** The synthesis of ZnS nanorods with an average of 8–10 nm in length and 1–2 nm in diameter has been reported previously[48,49], which we followed with some modifications. To briefly describe, the precursor Zinc-ethylxanthate ($Zn(SSCOC_2H_5)_2$) was prepared by dissolving 3.00 g of potassium ethyl xanthogenate and separately 2.05 g of zinc acetate dihydrate in water. The solutions were mixed together and zinc-ethylxanthate salt precipitated out. The salt was washed with water, filtered and dried in vacuum. ODA-coated ZnS nanoparticles were prepared using 0.08 g of as prepared zinc-ethylxanthate, which was dissolved in 1.53 g of molten ODA. ODA was initially exposed to $CO_2$ for 30 mins to form OAOC, zinc-ethylxanthate was added to molten OAOC, under Ar at 100 °C. The nanorod synthesis was carried out in two-steps, first at 105 °C for 5 min and then an additional 10 min at 130 °C. The ODA-coated ZnS nanoparticles were collected by flocculating the sample with ethanol, separated by centrifugation, redispersed in chloroform and drying in vacuum.

**Water Treatment.** The polar surfaces of the nanorods are relatively more reactive than the mixed Zn-S surface, and can trigger formation of new phases as well as precipitation of heavy cations present in an aqueous system[30]. An identical treatment protocol for ZnS nanorods[30] in aqueous solutions containing Fe(III) and As(III) was followed here with an idea of using the Zn-only ends as INAS facilitating Tooeleite nucleation within the overall neutral aqueous media. Three different water samples were targeted here: 1) As(III) contaminated ground water from Baruipur area of West Bengal, India (NW-Kolkata: NW denotes Natural Water); 2) a laboratory-made aqueous solution of As(III) and Fe(III) cations in concentrations matching the concentrations of sample 1 (AF-Lab: 'AF' denotes As and Fe); and 3) natural water similar to Baccu Locci mine area of Sardinia, Italy. Sample 3 was prepared using natural water from Iglesias, Sardinia, Italy, which had an identical aquatic composition as in Baccu Locci but without any arsenic, which was later spiked by As(III) in the laboratory with an amount equivalent to the concentration in Baccu Locci water (NW-Sardinia). All these details of the samples are tabulated in Supplementary Table S1 available online. For laboratory experiments, each 10 ml water sample was treated with 0.97 mg of solid containing ODA capped-ZnS nanorods; however, several scaled-up experiments showed that as little as 5 mg of the ZnS-ODA solid is sufficient for the decontamination of 1 litre of water. All treatments (10 ml or 1 L) were carried out at 70 °C for 3 hours, the resultant solutions were filtered using Whatman 42 filter paper (~2.5 μm pore size). In all cases, the corresponding filtrates and residues (whenever available) were collected for quantitative analysis. All of the results presented on the Iglesius water artificially spiked by As(III), are from scaled-up treatments. To ensure the recovery of residue mass, sufficient for different experimental analysis, the natural contaminated water was pre-concentrated significantly by prolonged water evaporation under continuous Argon gas flow (to nullify oxidation of As(III) to As(V)) e.g., the natural water sample from Baruipur, Kolkata was concentrated by 1000–2000 times.

**Characterization.** The chemical analysis of the water samples, both before and after treatment, and of the residue were carried out by ICP-OES (Perkin-Elmer Optima 2100 DV), ICP-OES (Shimadzu ICPE-9000), and ICP-MS instruments (ICP-MS, Elan DRC-e, Perkin-Elmer). The bulk structure of the residues were analyzed using powder X-ray diffraction (PXRD) experiments, which were carried out in the BL-18B Indian beamline of Photon Factory, Japan, and by 2-dimentional X-ray diffraction (2D-XRD) in ID09 beamline at the ESRF, France. The integrated 2D-XRD patterns were interpreted with a search-match algorithm through the ICSD database, and the presence of Tooeleite was revealed. A Le Bail fit using the monoclinic cell of Tooeleite allowed for lattice parameter extraction for the different samples. The quality of the AF-Lab_ZnS-NR pattern allowed for a structural determination, by the charge-flipping algorithm approach and Rietveld refinement, using the JANA2006 and superflip software[50,51]. First, a background profile fitting and then lattice parameters were refined using the Le Bail method. A reliable fit was obtained using anisotropic broadening modelling, which resulted in the significant broadening of the (001) peak and its harmonics. The extracted intensity was used for the *ab-initio* structure solution, which provided the positions of As and Fe and confirmed the *C2/m* symmetry. The oxygen coordinating these cations were located by a Fourier difference calculation. The interlayer molecules were not located because Fourier maps indicated a significant disorder, as was also confirmed by the (00l) peak broadening. The atomic coordinates of the refined atoms and a plot of the observed, calculated and difference profile curves are shown in Supplementary Table S3 and Fig. S1 available online.

The local structure of the solid was probed by X-ray absorption spectroscopy (XAS) at the Fe and As *K*-edges in the XAFS beamline of Elettra Synchroton Center, Italy. XAS data analysis was performed by looking at the near-edge (XANES) region (see Supplementary Fig. S2(b) for Fe *K*-edge XANES, available online) and compared with spectra of reference material, and fitting of the extended (EXAFS) signal was carried out within a multi-shell model using a local atomic cluster based on the Tooeleite crystallographic structure (ICSD code number 156179), as shown in Supplementary Fig. S2(d). The structural parameters of all of the samples obtained from XAS experiments (Table 1) matches well with the As and Fe local environment of the Tooeleite structure. X-ray photoelectron spectroscopy (XPS) experiments were carried out in the Materials Science beamline of Elettra Synchroton Center, Italy, and in an Omicron photoemission spectrometer equipped with EA-125 analyser and Mg $K_\alpha$ x-ray source. Supplementary Fig. S2(a,c) show the Fe $2p$ and As $3d$ core level spectra, respectively, from three solid residue samples, and confirm the presence of $Fe^{3+}$ and $As^{3+}$ ions in them. The morphology, composition and structure of nanorods and the solid residue were probed in a JEOL-JEM 2100F electron microscope using a 200 kV electron source ultrahigh-resolution transmission electron microscope (UHR FEG-TEM).

The electronic structure studies for all molecules and related species were performed using density functional theory (DFT), as implemented in the Gaussian 09 suite of programs[52]. Studies were performed using the B3LYP[53–55] hybrid functional with the 6–31G** basis set for all atoms. For any reaction Reactants → Products, the enthalpy of formation is calculated as:

$$\Delta H_{reaction} = \sum_{products} \Delta H - \sum_{reactants} \Delta H \tag{7}$$

The stable structures of all isolated molecules and complexes were determined by full geometry optimization in the gas-phase, and consequent harmonic frequency calculations were performed to ascertain the stationary points. To obtain the stable structures, a double-check was carried out using the PBE[56] functional along with a double zeta polarized (DZP) basis set employing the Quantumwise software[57–59]. The Fe complexes studied here are all low-spin, as there are even numbers of Fe-ions (experimentally also no net moment was found in residues). However, high-spin cases have also been looked into, theoretically. Though the absolute values of energies change for high spin complexes, all important results remain unaltered. The growth of the As(III)-Fe(III) template as in Tooeleite in all directions, predicted by our calculation, is schematically presented in Supplementary Fig. S3 available online.

## References

1. Ravenscroft, P., Brammer, H. & Richards K. *Arsenic Pollution A Global Synthesis* (Wiley-Blaclwell, 2009).
2. Kitchin, K. T. *Biological Chemistry Of Arsenic, Antimony And Bismuth* 1stEdn (Ed. Sun, H.) (Wiley, 2011).
3. *Arsenic Poisoning: A medical Dictionary, Bibliography, And Annotated Research Guide To Internet References* (Eds Parker, J. N. & Parker, P. M.) (ICON Health Publications, 2004).
4. *Fifth International Conference On Arsenic Exposure And Health Effects* (Eds Chappell, W. R., Abernathy, C. O. & Calderon, R. L.) (Elsevier, 2002).
5. Fendorf, S., Michael, H. A. & van Geen, A. Spatial and temporal variations of groundwater arsenic in south and southeast Asia. *Science* **328,** 1123–1127 (2010).
6. Nordstrom, D. K. Worldwide occurrences of arsenic in ground water. *Science* **296,** 2143–2144 (2002).
7. Rodríguez-Lado, L. *et al.* Groundwater arsenic contamination throughout China. *Science* **341,** 866–868 (2013).
8. Korte, N. E. & Fernando, Q. A review of arsenic (III) in groundwater. *Crit. Rev. Env. Contr.* **21,** 1–39 (1991).
9. U. S. Environmental Protection Agency, *Technologies and costs for removal of arsenic from drinking water.* (2000) Available at: http://water.epa.gov/drink/info/arsenic/upload/2005_11_10_arsenic_treatments_and_costs.pdf. (Accessed: 1st November, 2012).
10. Hao, L., Zheng, T., Jiang, J., Hu, Q., Li, X. & Wang, P. Removal of As(III) from water using modified jute fibres as a hybrid adsorbent. *RSC Adv.* **5,** 10723–10732 (2015).
11. Abdollahi, M., Zeinali, S., Nasirimoghaddam, S. & Sabbaghi, S. Effective removal of As (III) from drinking water samples by chitosan-coated magnetic nanoparticles. *Desalin. Water Treat.* **56,** 2092–2104 (2015).
12. Tandon, P. K., Shukla, R. C. & Singh, S. B. Removal of Arsenic(III) from water with clay-supported zerovalent iron nanoparticles synthesized with the help of tea liquor. *Ind. Eng. Chem. Res.* **52,** 10052–10058 (2013).
13. Monique, B. & Frimmel, F. H. Arsenic – a review. part I: occurrence, toxicity, speciation, mobility. *Actahydrochim. hydrobiol.* **31,** 9–18 (2003).
14. Gallegos-Garcia, M., Ramírez-Muñiz, K. & Song, S. Arsenic removal from water by adsorption using iron oxide minerals as adsorbents: a review. *Miner. Process. Extr. M* **33,** 301–315 (2012).
15. Mohan, D. & Pittman, C. U. Arsenic removal from water/wastewater using adsorbents-a critical review. *J. Hazard. Mater.* **142,** 1–53 (2007).
16. Höll, W. H. Mechanisms of arsenic removal from water. *Environ. Geochem. Health* **32,** 287–290 (2010).
17. Monique, B. & Frimmel, F. H. Arsenic - a review. part II: oxidation of arsenic and its removal in water treatment. *Actahydrochim. hydrobiol.* **31,** 97–107 (2003).
18. Choong, T. S. Y., Chuah, T. G., Robiah, Y., Gregory Koay, F. L. & Azni, I. Arsenic toxicity, health hazards and removal techniques from water: an overview. *Desalination* **217,** 139–166 (2007).
19. Chakraborti, D. *et al.* Status of groundwater arsenic contamination in the state of west bengal, india: a 20-year study report. *Mol. Nutr. Food Res.* **53,** 542–551 (2009).
20. Central Ground Water Board, Ministry of Water Resources, Government of India, *Arsenic in Ground Water In India.* (2009) Available at: http://cgwb.gov.in/documents/Bhujal_News_24_2.pdf. (Accessed: 10th December, 2012).
21. Times of India, *Arsenic in groundwater impacts 7 crore lives: Panel.* (2014) Available at: http://timesofindia.indiatimes.com/india/Arsenic-in-groundwater-impacts-7-crore-lives-Panel/articleshow/45486985.cms. (Accessed: 12th December, 2014).
22. Clancy, T. M., Hayes, K. F. & Raskin, L. Arsenic waste management: a critical review of testing and disposal of arsenic-bearing solid wastes generated during arsenic removal from drinking water. *Environ. Sci. Technol.* **47,** 10799–10812 (2013).
23. Cesbron, F. P. & Williams, S. A. Tooeleite, a new mineral from the U.S. mine, Tooele county, Utah. *Mineral. Mag.* **56,** 71–73 (1992).
24. Morin, G., Rousse, G. & Elkaim, E. Crystal structure of tooeleite, $Fe_6(AsO_3)_4SO_4(OH)_4 \cdot 4H_2O$, a new iron arseniteoxyhydroxysulfate mineral relevant to acid mine drainage. *Am. Mineral.* **92,** 193–197 (2007).
25. Hudson-Edwards, K. & Santini, J. Arsenic-microbe-mineral interactions in mining-affected environments. *Minerals* **3,** 337–351 (2013).
26. Morin, G. *et al.* Bacterial formation of tooeleite and mixed arsenic(III) or arsenic(V) - iron(III) gels in the carnoulès acid mine drainage, france. a xanes, xrd, and sem study. *Environ. Sci. Technol.* **37,** 1705–1712 (2003).
27. Xu, Y., Yang, M., Yao, T. & Xiong, H. Isolation, identification and arsenic-resistance of *Acidithiobacillusferrooxidans* HX3 producing schwertmannite. *J. Environ. Sci.* **26,** 7, 1463–1470 (2013).

28. Hamad, S., Cristol, S. & Catlow, C. R. A. Surface structures and crystal morphology of ZnS: computational study. *J. Phys. Chem. B* **106,** 11002–11008 (2002).
29. Balantseva, E., Berlier, G., Camino, B., Lessio, M. & Ferrari, A. M. Surface properties of ZnS nanoparticles: a combined DFT and experimental study. *J. Phys. Chem. C* **118,** 23853–23862 (2014).
30. Malakar, A., Das, B., Sengupta, S., Acharya, S. & Ray, S. ZnS nanorod as an efficient heavy metal ion extractor from water. *Journal of Water Process Engineering* **3,** 74–81 (2014).
31. World Health Organisation, *Guidelines for Drinking-water Quality.* (2011) Available at: http://whqlibdoc.who.int/publications/2011/9789241548151_eng.pdf. (Accessed: 5th December, 2013).
32. Chatterjee, A. *et al.* Arsenic in ground water in six districts of west bengal, india: the biggest arsenic calamity in the world part I. arsenic species in drinking water and urine of the affected people. *Analyst* **120,** 643–650 (1995).
33. Das, D., Chatterjee, A., Mandal, B. K., Samanta, G. & Chakraborti, D. Arsenic in ground water in six districts of west bengal, india: the biggest arsenic calamity in the world. part 2. arsenic concentration in drinking water, hair, nails, urine, skin-scale and liver tissue (biopsy) of the affected people. *Analyst* **120,** 917–924 (1995).
34. Das, D. *et al.* Arsenic in groundwater in six districts of west bengal, india. *Environ. Geochem. Hlth.* **18,** 5–15 (1996).
35. Gault, A. G. *et al.* in *Plasma Source Mass Spectrometry Applications And Emerging Technologies* (Eds Holland, G. & Bandura, D.) 112–126 (Royal Society of Chemistry, 2003).
36. Harvey, C. F. *et al.* Arsenic mobility and groundwater extraction in bangladesh. *Science* **298,** 1602–1606 (2002).
37. Duarte, A. A. L. S., Cardoso, S. J. A. & Alçada, A. J. Emerging and innovative techniques for arsenic removal applied to a small water supply system. *Sustainability* **1,** 1288–1304 (2009).
38. Lin, T.-F. & Wu, J.-K. Adsorption of arsenite and arsenate within activated alumina grains: equilibrium and kinetics. *Water Res.* **35,** 2049–2057 (2001).
39. Hao, J. M., Han, M. J. & Meng, X. G. Preparation and evaluation of thiol-functionalized activated alumina for arsenite removal from water. *J. Hazar. Mater.* **167,** 1215–1221 (2009).
40. Nishimura, T. & Robins, R. G. Confirmation that tooeleite is a ferric arsenite sulfate hydrate, and is relevant to arsenic stabilization. *Miner. Eng.* **21,** 246–251 (2008).
41. Opio, F. K. Thesis from Queen's University, Kingston, Ontario, Canada (2013). Available at: https://qspace.library.queensu.ca/bitstream/1974/7798/1/Opio_Faith_K_201301_PhD.pdf. (Accessed: 12th July, 2013).
42. Rigoldi, A. Thesis from University of Cagliari, Cagliari, Italy (2010). Available at: http://veprints.unica.it/628/1/PhD_Americo_Rigoldi.pdf. (Accessed: 27th September, 2012).
43. Krylova, V. & Andrulevičius, M. Optical, xps and xrd studies of semiconducting copper sulfide layers on a polyamide film. *Int. J. Photoenergy* **2009,** 1–8 (2009).
44. Montibon, E., Järnström, L. & Lestelius, M. Characterization of poly (3,4-ethylenedioxythiophene)/poly(styrene sulfonate) (PEDOT:PSS) adsorption on cellulosic materials. *Cellulose* **16,** 807–815 (2009).
45. Fantauzzi, M., Atzei, D., Elsener, B., Lattanzi, P. & Rossi, A. Xps and xaes analysis of copper, arsenic and sulfur chemical state in enargites. *Surf. Interface Anal.* **38,** 922–930 (2006).
46. Laajalehto, K., Kartio, I. & Nowak, P. Xps study of clean metal sulfide surfaces. *Appl. Surf. Sci.* **81,** 11–15 (1994).
47. Farrell, J. & Chaudhary, B. K. Understanding arsenate reaction kinetics with ferric hydroxides. *Environ. Sci. Technol.* **47,** 8342–8347 (2013).
48. Belman, N. *et al.* Hierarchical superstructure of alkylamine-coated ZnS nanoparticle assemblies. *Phys. Chem. Chem. Phys.* **13,** 4974–4979 (2011).
49. Belman, N. *et al.* Hierarchical assembly of ultranarrow alkylamine-coated ZnS nanorods: a synchrotron surface x-ray diffraction study. *Nano Lett.* **8,** 3858–3864 (2008).
50. Petříček, V., Dušek, M. & Palatinus, L. Crystallographic computing system Jana2006: general features. *Z Kristall* **229,** 345–352 (2014).
51. Palatinus, L. & Chapuis, G. Superflip - a computer program for the solution of crystal structures by charge flipping in arbitrary dimensions. *J. Appl. Cryst.* **40,** 786–790 (2007).
52. Gaussian 09, revision A.02, (2009). Gaussian series of program that provides state-of-the-art capabilities for electronic structure modeling. Gaussian, Inc., Wallingford CT. URL http://www.gaussian.com/.
53. Becke, A. D. Density-functional exchange-energy approximation with correct asymptotic-behavior. *Phys. Rev. A* **38,** 3098–3100 (1988).
54. Lee, C., Yang, W. & Parr, R. G. Development of the Colle-Salvetti correlation-energy formula into a functional of the electron density. *Phys. Rev. B* **37,** 785–789 (1988).
55. Vosko, S. H., Wilk, L. & Nusair, M. Accurate spin-dependent electron liquid correlation energies for local spin density calculations: a critical analysis. *Can. J. Phys.* **58,** 1200–1211 (1980).
56. Perdew, J. P., Burke, K. & Ernzerhof, M. Generalized gradient approximation made simple. *Phys. Rev. Lett.* **77,** 3865 (1996).
57. Atomistix tool kit version 12.8, (2012). A powerful software for DFT-based calculation for a large number of atoms. QuantumWise A/S, Copenhagen, Denmark. URL http://quantumwise.com/.
58. Brandbyge, M., Mozos, J. L., Ordejón, P., Taylor, J. & Stokbro, K. Density-functional method for nonequilibrium electron transport. *Phys. Rev. B* **65,** 165401 (2002).
59. Soler J. M. *et al.* The SIESTA method for ab initio order-N materials simulation. *J. Phys-Condens. Mat.* **14,** 2745 (2002).

### Acknowledgements
AM thanks CSIR, India for fellowship, BD thanks DST, India for funds (project no. SR/FT/CS-119/2010). SR thanks CSIR, India for funding (project no. (1269)/13/EMR-II). SR also thanks DST, India for funding (project no. WTI/2K15/74). Authors thank Dr. F. Podda for help with ICP-MS analysis, Dr. A. Ardu for TEM imaging, Prof P. Lattanzi coordinator of project ARSELLA-INNOVA RE, which funded experiments in Sardinia, Italy. Authors also thank ICTP for funding experiments at XAFS beamline, Elettra, DST, India and Indo-Italian POC for the financial support in performing experiments in Materials Science beamline, Elettra and Saha Institute of Nuclear Physics, India for facilitating experiments at Indian Beamline, Photon Factory, KEK, Japan.

### Author Contributions
The work presented here is an outcome of discussions and interpretation of data among the authors. A.M. synthesized the nanorods and also carried out most of the experiments, while S.I. helped in synthesis and characterization. B.D. has carried out all the theoretical works. C.M., A.I. and G.A. have carried out the XAFS experiments. M.M. performed the 2D-XRD and analysis, while Y.V.K. partially carried out ICP experiments. G.D.G. helped in all the experiments in Sardinia and also carried out HRTEM experiments. S.A. helped in nanorod synthesis. S.R. wrote the paper and all the authors provided feedback to the manuscript. S.R. conceived the project and supervised the research.

### Additional Information

**Supplementary information** accompanies this paper at http://www.nature.com/srep

**Competing financial interests:** An Indian patent application (Ref. No. 822/KOL/2015) for the technology has also been filed. We have also attempted to build a pilot plant on Sardinia Island, Italy, under the ARSELLA project, funded by the Regional Administration of Sardinia and University of Cagliari. The value of these may get affected by the publication of this manuscript.

**How to cite this article**: Malakar, A. *et al.* Efficient artificial mineralization route to decontaminate Arsenic(III) polluted water - the Tooeleite Way. *Sci. Rep.* **6**, 26031; doi: 10.1038/srep26031 (2016).

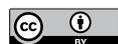 This work is licensed under a Creative Commons Attribution 4.0 International License. The images or other third party material in this article are included in the article's Creative Commons license, unless indicated otherwise in the credit line; if the material is not included under the Creative Commons license, users will need to obtain permission from the license holder to reproduce the material. To view a copy of this license, visit http://creativecommons.org/licenses/by/4.0/